\documentclass[preprint,12pt]{elsarticle}

\usepackage{amssymb}
\usepackage{amsmath}
\usepackage{xcolor}
\usepackage{float}

\usepackage{url}  

\journal{Physica A}
\begin{document}

\begin{frontmatter}



\title{Physical models of traffic safety at crossings}


\author[dlr]{Andreas Leich}
\author[dlr]{Ronald Nippold}
\author[uc]{Andreas Schadschneider}
\author[dlr,tub]{Peter Wagner}

\affiliation[dlr]{organization={Institute of Transport Systems, German Aerospace Centre},
            city={Berlin},
            postcode={12489}, 
            country={Germany}}
			
\affiliation[uc]{organization={Institute of Theoretical Physics, University of Cologne},
            city={Cologne},
            postcode={50937}, 
            country={Germany}}

\affiliation[tub]{organization={Institute of Land and Sea Traffic, Technical University of Berlin},
	city={Berlin},
	postcode={10553}, 
	country={Germany}}

\begin{abstract}
Traffic safety at intersections is studied quantitatively using methods from Statistical Mechanics on the basis of simple microscopic traffic flow models. In order to determine a relationship between traffic flow and the number of crashes, the modelling focus is on the building block of any road network, namely the crossing of two streams. In this paper, it is shown that the number of crossing conflicts is proportional to the product of the two traffic flows from which a simple model is developed. This model substantiates known empirical findings. Since real crash data are obtained by an involved process from such building blocks, there is a difference between the theoretical and empirical results. This process is modelled here as well and narrows the gap between theory and observation. 
\end{abstract}


\begin{highlights}
\item By using physical models for the conflicts at two crossing streams of vehicles, the relationship between the traffic flows of two crossing streams and the number of conflicts can be mapped. 
\item For small flows, a simple analytical model is in line with the simulation results, while it seems different from empirical results. 
\item While this might be due to human factors, this work instead models the step from the simulation to reality as well. This leads to a narrowing of the gap between simulation results and empirical results.
\end{highlights}


\end{frontmatter}



\section{Introduction}
\label{sect:intro}

The relation between the number of crashes and the number of vehicles at some location of a road network is fundamental for road safety research. It has been studied extensively in the traffic engineering literature \cite{Veh1937, Hauer2004, LordMannering2010, KononovEtAl2011, Hoye:Hesjevoll:2020}, mostly by analyzing empirical data with regression models that lack a physical or mechanistic model behind.

In addition, some theoretical studies, often by physicists, have investigated the occurrence of accidents or dangerous situations in specific scenarios. However, almost none of these investigations has done a comparison with empirical data. They are often theoretical approaches that focus on rear end collisions due to violations of the safety distance. 

One of the first studies \cite{Nagatani:1993} considered the effect of a block, caused by a crash, in a network. Later, a crash model was introduced \cite{Boccaraetal:1997,Huang:1998} that is based on simple criteria derived from the motion rules of the model (often variants of the Nagel-Schreckenberg model \cite{Nagel1992}). However, the ensuing crashes are never executed explicitly. Instead, the simulation is continued as without an accident. Therefore, the terminology “accident” has been replaced by “dangerous situations” in later works. Such a procedure will lead to correlations as it becomes rather likely that the criteria for a potential crash are again fulfilled immediately after such a “dangerous situation”. This leads to an overestimation of the number of crashes in such a setting, especially since most works considered periodic boundary conditions.

Here we attempt to bridge the gap between theoretical and empirical studies using concepts from Statistical Mechanics and simple microscopic traffic flow models to  obtain quantitative results that can be compared with empirical data. This delivers a deeper insight into the emergence of road crashes and the process of measuring crash rates.

The approach here is restricted to conflicts between two vehicles, which is the overwhelming majority of all crashes (more than 90\,\% in urban areas), and to the crossing of two streams of vehicles. In urban areas, more than 50\,\% of all crashes are of this type. 

\subsection{Modelling Crashes using Cellular Automata}
\label{sect:intro:models}

Cellular automata (CA) have been used to simulate traffic flow since the early 1990s \cite{Nagel1992}, from simple set-ups where all vehicles run in a circle to fairly complex network geometries, and to the simulation of different modes of traffic (cars, bicycles, pedestrians). It is not easy to give a concise review here, but see \cite{Chowdhury-Review:2000} (updated in \cite{SchadschneiderChNi}) for some more or less complete overview of early work. Road traffic safety simulation has received less attention in both the physics and transportation safety research communities. For a few examples see \cite{Boccaraetal:1997, Ito:Nishinari:2014, Marzoug:etal:2018, Marzoug:etal:2021} and \cite{Janvier:Saunier:2020} for the traffic safety community. In physics, especially the TASEP (Totally Asymmetric Simple Exclusion Process) and related models have been used to study crossings, however usually focusing on the traffic flow aspect and not on safety \cite{Akriti:Gupta:2021}. The reasons for this are that those models are constructed crash-free and so it is not clear whether any concept of traffic safety might apply here. Furthermore, crashes are rare events. Therefore, it is difficult to pin down detailed reasons for each crash that then can be modelled in a physically appealing manner. 

Of special interest is the relationship between the  number of crashes $N$ (per time-interval) and the traffic flow $Q$, since this is the most important factor that determines the number of crashes. In the following, capital letters are used to name the variables as measured in traffic safety studies, which is in most cases the long-term sum of the number of crashes $N$ (typically aggregated over one or more years and across different crash-types), and the so called Average Daily Traffic $Q$ (often named ADT). An ADT-flow is the average of the daily traffic of many days; in the following, to avoid confusion, $Q$ is often named ADT-flow. 

The simulation models below work on a shorter time-scale and in a fully controllable environment. This produces similar variables $n$ (the number of conflicts) and traffic flows $q_1, q_2$ for a crossing. It is assumed that the relationship found between $n$ and $q_1, q_2$ can be transferred to the macroscopic variables $N$ and $Q_1$, $Q_2$, or sometimes even to the total flow $Q := Q_1 + Q_2$. This transformation is non-trivial, the section \ref{sect:fromMicro2Macro} will deal with it. 

\subsection{Empirical Examination of Crashes vs.\ Traffic Flow}
\label{sect:intro:empirics}

The traffic safety community has come up with a lot of empirical work on this topic, without reaching a clear conclusion. From a recent meta-analysis \cite{Hoye:Hesjevoll:2020}, which assumed $N \propto Q^\beta$, at least an idea for the dependency of the number of crashes $N$ on the ADT-flow $Q$ may be found. Here, $Q$ is the total ADT-flow observed at a road or an intersection. In this case, the average of the studies included in the meta-study leads to exponents $\beta = 0.522$ for single vehicle crashes, and $\beta=1.210$ for multi-vehicle crashes. Other work dealing with intersections claim that $N \propto Q_1^{\beta_1} Q_2^{\beta_2}$ (here $Q_1, Q_2$ are the ADT-flows of the crossing roads, one is the major flow, one is the minor flow), where $\beta_i<1$ \cite{HSM2010}. Note, however, that the power-law is at best an approximation to small demand. For a larger demand, a kind of saturation might occur \cite{KononovEtAl2011,WaPetAl:2021}.

In general, traffic safety is modelled by so called safety performance functions where it is assumed that the number of crashes $N$ depends on the ADT-flows $Q_i$ and on other factors $x_i$ as follows \cite{Hauer2004,LordMannering2010,HughesEtAl2015,ManneringBook2018,Ambros2018}:
\begin{equation}
N = \beta_0 Q_1^{\beta_1} Q_2^{\beta_2} \exp \left( \sum_{i>2} \beta_i x_i  + \xi \right) \label{eq:glm}\!\!.
\end{equation}
Here, the $x_i$ are factors believed to influence $N$ as for instance the speed-limit, intersection organization and the like, and the $\beta_i$ are the coefficients that determine how strongly each $x_i$ influences the final outcome. In the case of the flows, they are the exponents with which the crash-rate grows as function of the $Q_i$. As mentioned already, the traffic flows in this equation are the average daily traffic (ADT); these $Q_i$ are the result of a sum over complex traffic flow time-series $q_i(t)$ which are often not known exactly. In addition, in empirical work it is often difficult to come up with reasonable and trustworthy values of $Q_i$ at all the places where crashes occur. This is because, especially for small crash numbers, there are issues with under-sampling, and even with under-reporting. And finally, Eq.~(\ref{eq:glm}) is just a regression analysis where a detailed model in the physical or mechanical sense is missing.

The noise-term $\xi$ is constructed so that the distribution of $N$ follows a negative binomial distribution (NBD). The NBD is a generalization to the Poisson distribution (PD). Its variance is larger than that of the PD with a parameter $\gamma$ that describes how the variance $\sigma^2$ depends on the mean value $\mu$ of the distribution. This is also named over-dispersion, since the width of the crash-number distribution is larger than what could be expected from a Poisson distribution. While $\sigma^2 = \mu$ holds for PD, the NBD has
\begin{equation}
\sigma^2 = \mu + \gamma \mu^2 , \label{eq:var-vs-mu}
\end{equation}
i.\,e.~the variance increases quadratically with the mean-value. For $\gamma=0$, the NBD is a PD. Eq.~(\ref{eq:var-vs-mu}) is an exact relation that holds for any NBD distribution.

\begin{figure}[h]
\includegraphics[width=0.77\textwidth]{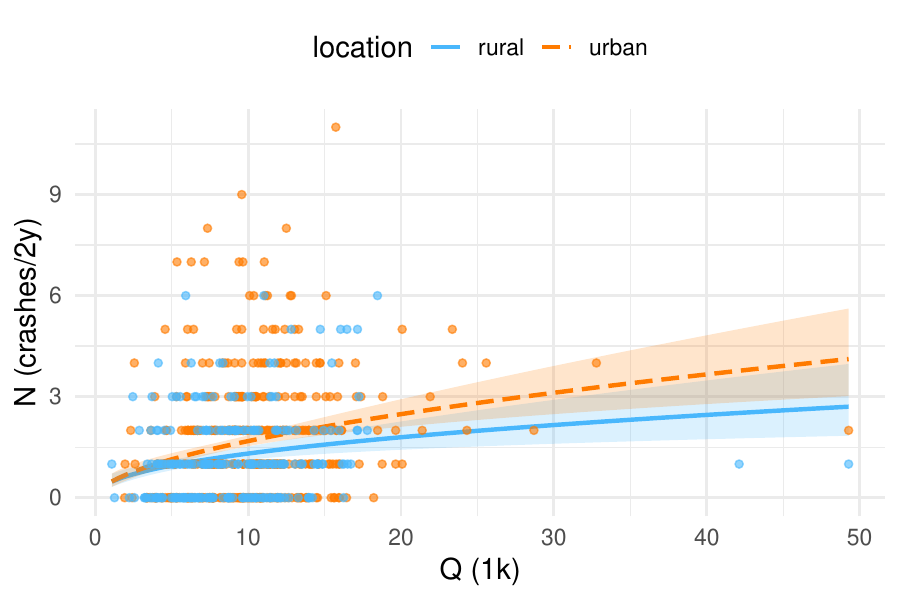}
\caption{Fit of a GL-model to crash data from German roundabouts. Plotted is the number of crashes versus the total flow $Q$ (sum over the flows of all approaches), both for roundabouts in urban areas and in rural areas (for more details, see \cite{Leich:2022}). \label{fig:RA}}
\end{figure}

The particular form in Eq.~(\ref{eq:glm}) is used since it is possible to fit real data to the mean of $\log N$ by a generalized linear model (GLM), and rely on the strong mathematical and statistical foundation of GLM's. The downside is that it is often difficult to check if this form is really consistent with the data. Admittedly, doing so is not trivial, as Fig.~\ref{fig:RA} demonstrates (included here only as an illustration). Fig.~\ref{fig:RA} displays the number of crashes versus the ADT-flow $Q$ for a subset of roundabouts in Germany, together with a fitted GLM. It can be seen that parameter values of the model fits are strongly influenced by a few data-points for large $Q$, and that the bulk of the data at smaller ADT-flow could be better described by a model where $N \propto Q$. 

\subsection{Outline}
\label{sect:intro:contribution}
 
\begin{figure}[h]
\includegraphics[width=0.61\textwidth]{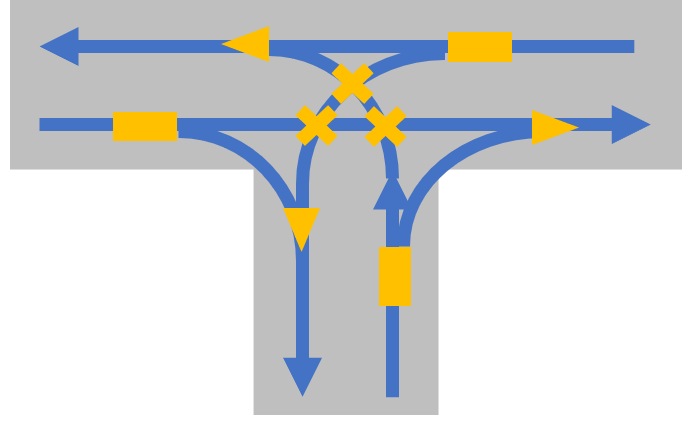}
\caption{Scheme of the conflict areas of a 3-leg or 
T-intersection. The crosses show the three crossing conflicts, the three rectangles rear-end (sometimes named diverging conflicts), and the triangles could be both (merging conflicts). Such a T-intersection has six possible flows. \label{fig:iX}} 
\end{figure}

This work is restricted to one of the building blocks of a road traffic system: the crossing of two traffic streams. Each real-life intersection is constructed of many conflict areas where two streams cross, see Figure~\ref{fig:iX} for a visualization. 

The simplest 3-leg intersection has three areas where streams cross, three areas where vehicles merge, and three areas where rear-end conflicts could occur, each with its own pair of demands (this counts only the conflicts between vehicles and ignores the ones between vehicles and pedestrians and is true for one-lane roads only). This work concentrates on just one conflict area (always named X in the following). With regard to traffic safety, this simplifies the matter greatly to essentially two types of conflicts: the ones that stem from the crossing itself, and the ones that are there due to the fact that a crossing is a capacity bottleneck which causes more braking compared to free flow, thus leading to rear-end conflicts in the approaching links. And it makes it treatable by simple simulation models, so that some complexity of the real-world can be dropped.

Note that empirical data are always a mixture of these and many more conflicts-types. However, since crashes are rare, it is in most cases not possible to discern the different conflict types and their dependency on traffic flow and other external circumstances, so they are lumped together in one crash number.

To connect models with empirical data, we concentrate on the dependency of the number of crashes $N$ on flow only. We focus on the pure mechanics, without considering the many human factors that may influence the crash probability. To do so, two carefully crafted CA models of a conflict area X are used as mechanistic models for the conflicts and their number, and the related number of crashes. 

\section{A short theory and two simulation models}

\subsection{Crossing and rear-end conflicts}

From the set-up above, two simple considerations might be put forward. Given are two streams of traffic objects with volumes (traffic flows) $q_1, q_2$ wanting to cross a conflict area of size $L_1 \times L_2$. The streams are assumed to be statistically independent, which means that each flow $q_i$ emits with a certain probability $p_i = q_i\Delta t$ a vehicle at random, where $\Delta t$ is the time-step size. Each vehicle has a length $\ell_i$ (they can be made different, but here it is assumed that they are equal for all of them), and that means that it needs the time $\tau_i = (L_i + \ell_i)/v_i$ to cross X, where $v_i$ is the speed right at X. The average time between two vehicles is the gross time headway $T_i = 1/q_i$. From the assumption of independence the following deduction holds for the probability $\Pi$ that these two end up being in conflict ($L_i = \ell_i = \ell$ has been set to simplify the resulting expressions):
\begin{equation}
\Pi_{\text{X}} = \frac{\tau_1}{T_1} \frac{\tau_2}{T_2} = 4\ell^2 \frac{q_1\,q_2}{v_1\,v_2} \label{eq:theory:X}
\end{equation}
Here, the subscript 'X' has been used to indicate crossing conflicts. The same can be done for the conflicts caused by rear-end interactions (subscript 'RE'):
\begin{equation}
\Pi_{\text{RE}} = \frac{\tau_1}{T_1} \frac{\tau_1}{T_1} + \frac{\tau_2}{T_2} \frac{\tau_2}{T_2} = 4 \ell^2 \left( \left( \frac{q_1}{v_1}\right )^2 + \left( \frac{q_2}{v_2}\right )^2 \right ) \label{eq:theory:RE}
\end{equation}

For completeness, the total number of conflicts may then be written as:
\begin{equation}
\Pi =  4\ell^2 \left( \frac{q_1\,q_2}{v_1\,v_2} + \left( \frac{q_1}{v_1}\right )^2 + \left( \frac{q_2}{v_2}\right )^2 \right ) \label{eq:theory}
\end{equation}

The notable property of these equations is that they connect the probability for conflict with two traffic flow parameters -- the speed and the flow. Clearly, an alternative formulation would be to use the densities $\rho_i = q_i/v_i$; this is not done here, since in traffic safety speeds are typically not available, except for freeway data. 

For an intersection, there is of course a complicated relationship between these four variables, which in most cases can only be mapped by simulations, especially for larger flows. In addition, it depends on the organization of the intersection, in this paper the symmetric “first come first serve” (FCFS) will be used, but others can be considered as well.

\subsection{From conflicts to crashes}

This approach might also be a good starting point for more detailed crash models. The simplest one, which is used here, states that the actual crash probability $\pi$ is simply a constant factor $\alpha$ to be multiplied with $\Pi$, where $\alpha$ is a very small number (e.\,g.~$\alpha = 10^{-3} \ldots 10^{-6}$), and it might be different for crossing as well as for rear-end conflicts:
\begin{equation}
\pi = \alpha \Pi \label{eq:Pipi}
\end{equation}

Note, too, that the traffic safety research community's predominant models for a whole intersection \cite{HSM2010, RetallackOstendorf2020} are not exactly the same as Eq.~(\ref{eq:theory}):
$$
N = \pi T_{\text{obs}} \propto Q_1^{\beta_1} Q_2^{\beta_2}
$$
Here, $T_{\text{obs}}$ is an observation period. To our knowledge, speeds have not been used in such models apart from a general speed-limit that may or may not be in place at an intersection. With the exception of freeway data, speeds are almost never available for intersections. 

More complicated dependencies of $\alpha$ on the traffic state might be considered, but this means to enter the realm of human factors. This will not be done here. Instead, we concentrate on a better understanding of the relationship between $\Pi, \pi$ and the traffic state described by $q_i, v_i$. If $\alpha$ is in fact a constant, then the number of crashes is directly proportional to the number of conflicts, and therefore it suffices to regard the latter ones instead of real crashes. From Eq.~(\ref{eq:theory}) it might be expected that $n \propto q_1 q_2 + q_1^2 + q_2^2$ is the right relationship between $n$ and $q$. As will be seen in the results section, this is only true for small flows, as long as the speed $v$ is mostly independent of the traffic flow $q$. For larger flows, the interaction between the objects gains influence and changes this simple picture. 

\subsection{Models}

Two models have been chosen here, the CA model with maximum speed $v_{\text{max}}=2$ (measured in sites/time-step, where the length of a site is the generalized vehicle length of $\lambda=7.5\,m$, and the TASEP (totally asymmetric simple exclusion process) with parallel update. The time-step $\Delta t$ has been chosen to be 1 second for the CA, and $\Delta t=0.5\,s$ for the TASEP, respectively, which leads to the same speed of 15\,m/s when expressed in real-world units. Each stream is a set of vehicles that are described by their position as well as their speed, $\{(x_i(t),v_i(t))\}_{i=1,\ldots,K(t)}$. The number of objects $K(t)$ in each stream is time-dependent. Open boundary conditions are used: Objects try to enter with a certain rate $p_i \Delta t$ and can enter only if there is a free space at the beginning of the link. Therefore, the flow that is finally achieved and that determines the number of crashes is always smaller or equal to $q_i \leq p_i \Delta t$. The downstream boundary is completely open without any restriction on leaving. This outflow is what is named flow $q_i$ in the following, and this is the variable that determines the conflict-rates in Eq.~(\ref{eq:theory}). The crossing site is located $v_{\text{max}}$ sites upstream of the last site, this helps to catch the interaction between vehicles approaching X and the ones leaving X. 

Choosing these two models has an advantage over more complex modelling approaches (like continuous models as those implemented in micro-simulation software): A conflict can be determined unambiguously and therefore can be counted. The speed and position update rules, which are repeated here for the sake of completeness ($g$: number of empty cells in front, $v, v', \hat v$: current, updated, and intermediate speeds, respectively, rand() a random-number in $(0,1]$, and $p_{\text{brake}}=0.25$ a randomization probability) are:
\begin{eqnarray}
&& \hat v = \min \{v + 1, v_{\text{max}}, g\}, \\
&& v'  = \left \{ \begin{array}{ll}
\max\{0, \hat v - 1 \} & \text{if} \,\, \text{rand()}<p_{\text{max}} \\
\hat v & \text{else} 
\end{array} \right . ,\\
&& x' = x + v'.
\end{eqnarray}
They have to be supplemented with rules that are applied to X, which will be stated below. 

The TASEP update rule is just the one of the CA with $v_{\text{max}}=1$. Some simulation experiments have also been performed with a modified randomization rule for the TASEP (where the probability to move after standstill increases with standing time), but they yield very similar results and so are only mentioned here to confirm the robustness of the results below.

The conflicts are handled as follows: The crossing area X has an own discrete variable, which can be either free, or a block to one of the two links. For the TASEP (starting from an unblocked state), in each time-step it is checked whether there is an object on the two sites upstream of X. If just one of these two sites is occupied, the other TASEP's site X is blocked. This blocking state is retained until the first object has cleared X, which needs at least two time-steps. If both sites upstream of X are occupied (no matter the speed of the two objects), then this is recorded as a conflict, and it is resolved by deciding randomly which object is allowed to run first. This conflict is counted, and the respective exit times are recorded as well -- they will be used in the subsequent analysis below. 

For the CA, the handling of conflict area X is more complicated: Here, all sites that can reach X in the next time-step are checked for the presence of objects and counted if at least one is present. If on one link the count is larger than zero and the other is zero, then the other link gets blocked. If there are objects on both links that can reach X in the next time-step, then it is decided randomly which one moves first towards X. And, as in the case of the TASEP, it is counted as a conflict. Again, the block is removed after the object has cleared X completely; this may happen in this case in one time-step.  
\begin{figure}[h]
\includegraphics[width=0.77\textwidth]{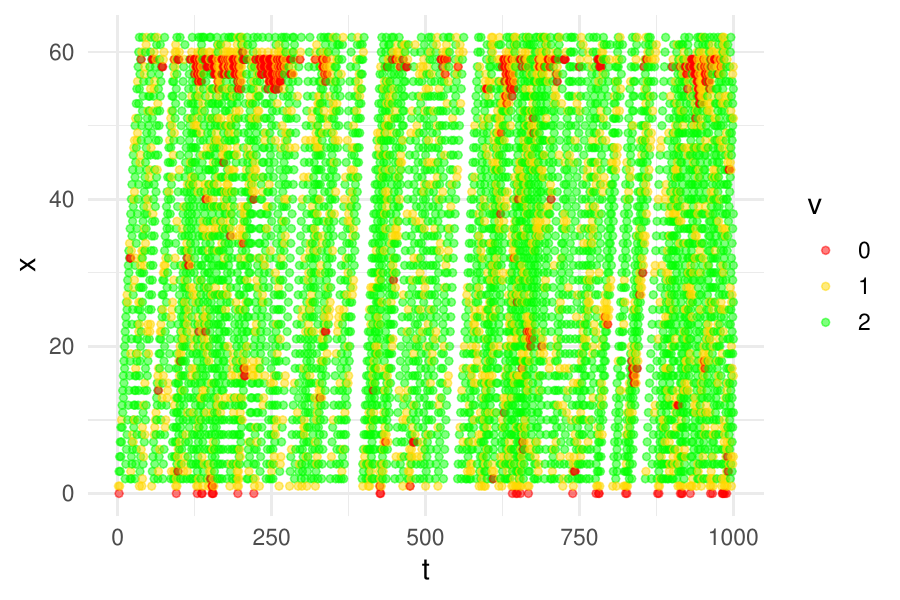}
\caption{Space-time diagram of one link of CA objects approaching conflict site X, which is located at site 60. Here, X corresponds to one of the crossing points in Figure~\ref{fig:iX}. Note that $v_{\text{max}}$ additional sites downstream of X have been included to catch the interaction between a vehicle approaching X and the one leaving X. Each point is colored according to the speed. \label{fig:stCA}}
\end{figure}

A typical space-time diagram for one link is displayed in Figure \ref{fig:stCA}. 

All simulations have been run for at least $10^5$ time-steps per traffic flow value. For small demand, typically more time-steps are needed to reach a reasonable number of events: In this case, the simulation is run until at least 100 conflicts had been recorded. 

The rear-end conflicts are counted differently. For the TASEP, any interaction where the following vehicle is blocked by a leading vehicle is counted as a conflict. For the CA model, all interactions where the speed was adjusted to the distance ahead, i.\,e.~where $\hat v = g$ had been applied, are counted as a conflict. This includes two standing objects, since in principle the follower could move and crash. 

Clearly, at least for the CA-model a kind of severity might be assigned to each conflict, depending on the speed-difference between them. Again, for the sake of simplicity, this work refrains from such additional complications.

\section{Results}

\subsection{Conflicts versus flows}

\begin{figure}[H]
\includegraphics[width=0.9\textwidth]{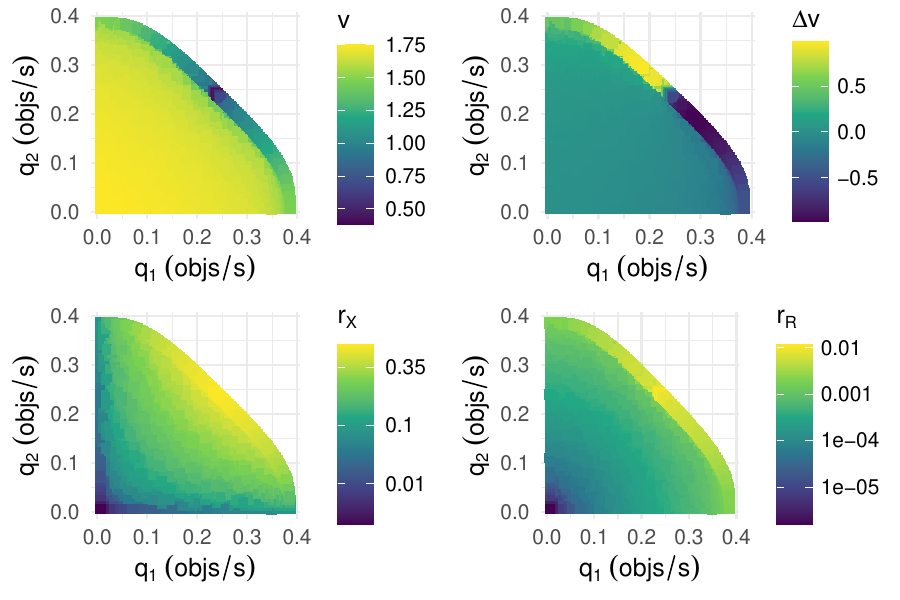}
\caption{The average speed $v$ (top left), the speed difference $\Delta v = v_1 - v_2$ (top right), and the two crash-rates $r_\text{X}$ (lower left) and $r_\text{RE}$ (lower right) as function of the two flows $q_1, q_2$ for the CA model. \label{fig:CA2D}}
\end{figure}

The result of the simulations are several functions $f(q_1,q_2)$ that characterize the system's response. Here, $f$ are chosen to be the average speed, the speed difference $\Delta v = v_1 - v_2$ between the two links, the number of crossing conflicts $n_\text{X}(q_1, q_2)$ and the number of rear-end conflicts $n_\text{RE}(q_1, q_2)$. Plots of those quantities are displayed in Figure \ref{fig:CA2D} for the CA model. In the following, a normalized conflict rate will be used: All numbers of conflicts are divided by the length of the time-interval within which they occurred, while the rear-end conflicts, in addition, are divided by the number of cells used in the simulation, therefore giving a rate per cell. The normalized conflict rates are named as  $r_\text{X}(q_1, q_2)$ and $r_\text{RE}(q_1, q_2)$, respectively, and the sum of both is the conflict rate $r(q_1, q_2)$.

\begin{figure}[H]
\includegraphics[width=0.85\textwidth]{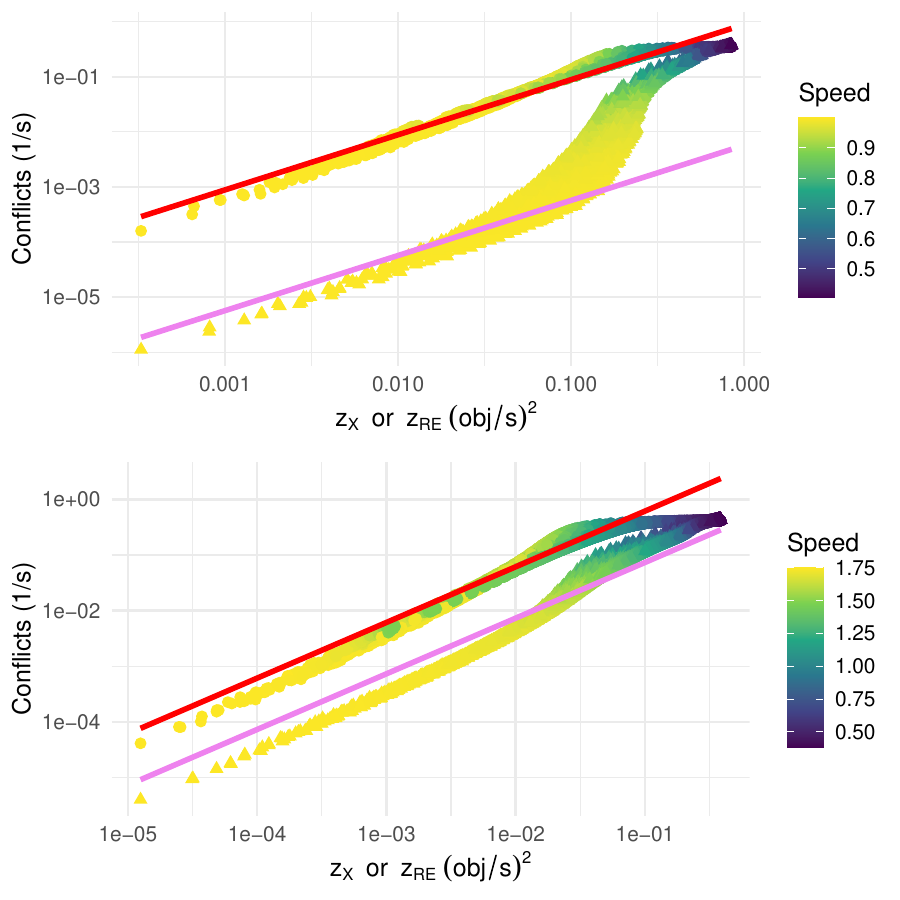}
\caption{The two rates $r_\text{X}$ (top cloud of points) and $r_\text{RE}$ (bottom cloud of points) are plotted versus the variables $z = z_\text{X}$ or $z_\text{RE}$, respectively. They follow the theoretical approach quite nicely. The color of the points indicates the speed, the red (for $z_\text{RE}$) and violet (for $z_\text{X}$) lines are linear fits. The figure on top is from the TASEP, the one on the bottom from the CA-model. \label{fig:r-vs-z}}
\end{figure}

In Eq.~(\ref{eq:theory}), conflict rates are simple functions of the traffic state $z$, where $z_\text{X} = q_1q_2/(v_1 v_2)$ for the crossing conflicts, and $z_\text{RE} = (q_1/v_1)^2 + (q_2/v_2)^2$ for the rear-end conflicts. Therefore, Figure ~\ref{fig:r-vs-z} displays the rates versus the two $z_x$ variables for the two simulation models, together with a linear fit of $r_x$ versus $z_x$ (the fit is done with weights of $1/r_x^2$ to favor small values of $z$). This shows that the considerations above are very well in line with the simulation results. 
For medium and large values of $z$, deviations can be seen: The number of crossing conflicts saturates, while the number of rear-end conflicts first grows stronger than in theory, but seems to saturate as well when approaching the capacity of the link.

\subsection{Empirical analyses and the Average Daily Traffic}
\label{sect:fromMicro2Macro}

The results so far are not completely in line with the empirical results. The main differences are: The exponents of the $N(Q)$ relationship in empirical data are smaller than in the simulation, and the distribution of crash numbers is different from the Poisson distributions obtained in the simulation. This could be the already mentioned negative binomial (NBD), or any other distribution with over-dispersion. Therefore, a closer look is needed how the empirical results are obtained, given the rarity of real crashes.

One data-point $(Q_k, N_k)$ in an empirical analysis is obtained e.\,g.~by the number of crashes $N_k$ over a certain time-period (typically of the order of months and years) at a certain place $k=1,\ldots,K$, while the corresponding ADT-value $Q_k$ in ideal cases is counted over the same period. More often, it is obtained by short-term counts extrapolated, or it is based on a travel demand model. Even in good conditions, each $Q_k$ is a sum over all the different traffic flows (enumerated by $\nu$) in this place (think of the six streams at the three-leg intersection above), and it averages over time $t$:
\begin{equation}
    Q_k = \frac{1}{T} \sum_{t,\mu} q_k^\nu (t)
\end{equation}

\begin{figure}[bh]
\includegraphics[width=0.85\textwidth]{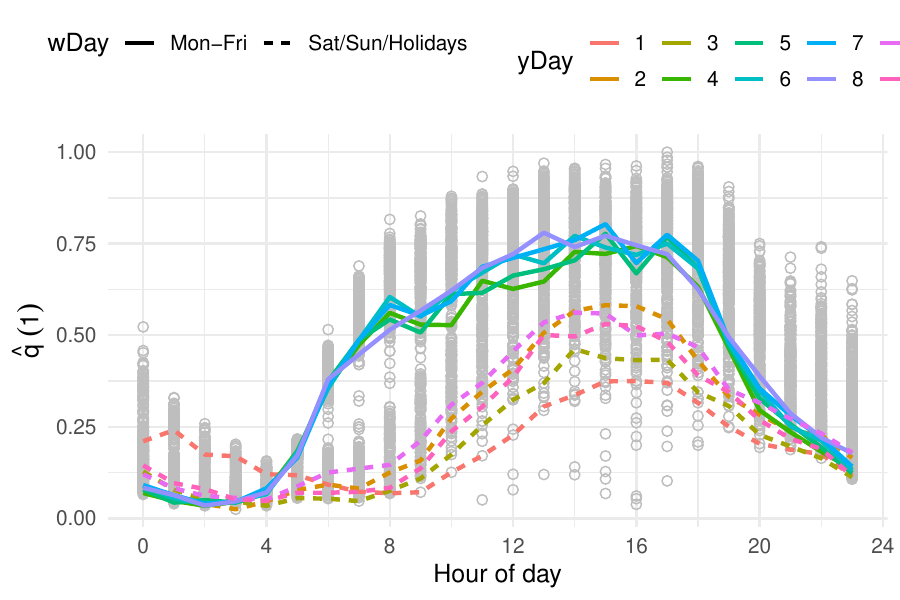}
\caption{Scaled flow values (gray values) of all flow values, and ten daily courses $\hat Q_i(t)$ (lines) for one typical loop detector in Berlin. \label{fig:TE370}}
\end{figure}

The daily curves $q_k^\nu (t)$ display a wide range of shapes, one example could be seen in Figure~\ref{fig:TE370} with data from one detector in Berlin, the data-set has been obtained from \cite{BerlinLoops}. 
Similar data-sources can be found elsewhere \cite{UTD19}. It is assumed that there will be no principal difference between the data from Berlin and data from other places. 

This sampling process can be modelled as well, with a few assumptions about the form of the various distributions needed:
\begin{itemize}
    \item The distribution of $q_k(t)$-values that are used to compute $Q_k$ is simply a uniform distribution in $[0,2\,Q_k]$.
    \item The distribution of the $Q_k$-values itself is again a uniform distribution, this time in the interval $10^3, 5\cdot 10^4$ vehicles/day.
    \item From this, artificial crashes are generated by setting a crash probability $p = \alpha q^\beta$, resulting in a reasonable number of collisions.
\end{itemize}

\begin{figure}[th]
\includegraphics[width=0.85\textwidth]{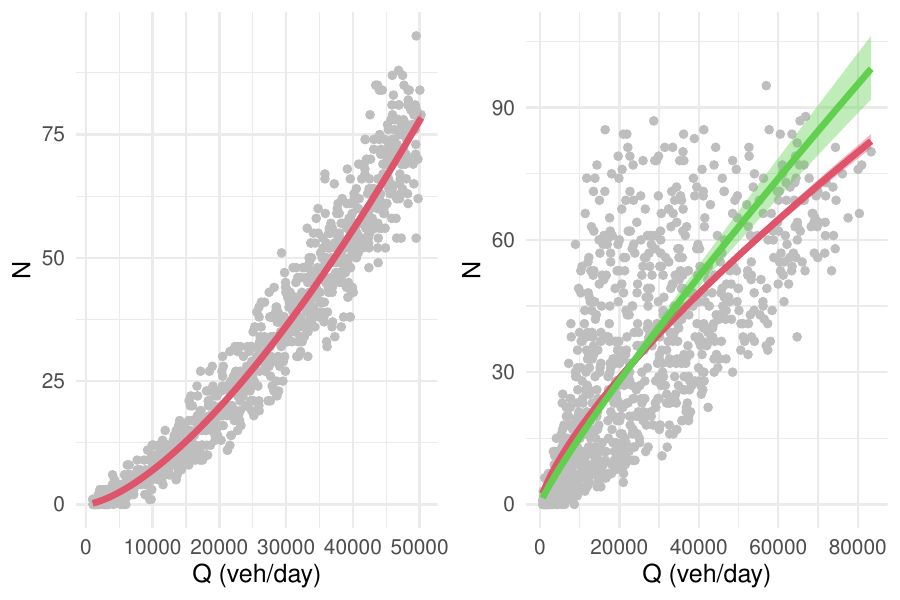}
\caption{Simulated measuring process. Left, without error, the red line is a fit to a model with $Q^\beta$ and a Poisson distribution. The plot on the right is the same crash data, in this case all $Q$-values have been shuffled as described in the text. Now, the exponent has changed. The red line is again a fit with $Q^\beta$, as is the green line. The red line is from a Poisson distribution, while the green is for a negative binomial one.   \label{fig:simEmpirics}}
\end{figure}

Doing this for a simulated for 1000 days and $K=1000$ places, the resulting $N(Q)$-curve still reproduces the exponent and the crash-number distribution is $P$ (see Fig.~\ref{fig:simEmpirics}, left plot). 
Several approaches have been tried, the by far the simplest one is to assume that the $Q_k$-data are strongly disturbed. By multiplying each $Q$-value with a number drawn from a uniform distribution in $[0.25,1.75]$ and then repeating the fitting process, the exponent changes, and the distribution switches to an over-dispersed one.  

The result of this procedure is shown in Fig.~\ref{fig:simEmpirics}. It is very likely that in the real world more complicated processes are in place that modify the Poisson distribution at the heart of this process, but for this paper the simplest possible explanation is used.

\section{Conclusions}

This research has tried to find a relationship between the traffic flow and the number of crashes at the crossing of two traffic streams. This is achieved by investigating the behavior of two simulation models and therefore yields a mechanism that can explain some of the empirical observations. It turned out to be important to make a distinction between the microscopic picture drawn by the simulation, which leads to a relationship between the traffic flows $q_i$ and the number of conflicts $n$, and the macroscopic picture from empirical results, which yields a similar relationship between the macroscopic ADT-flows $Q_i$ and the number of crashes $N$. 

The simulation results confirm a theoretical approach that shows that the number of conflicts behaves as simple functions of the traffic state, see Eq.~(\ref{eq:theory}). The main result is that especially for small flows, the number of conflicts $n$, and with it the number of crashes, should in fact be $n \propto q_1 q_2$ for crossing conflicts, and $n \propto q_1^2 + q_2^2$ for rear-end conflicts. For large flows, more complicated behavior becomes visible: For intermediate flows, the crash-rate seems to grow stronger than $n \propto q_1 q_2$, while at very large flows close to capacity, saturation becomes visible. The increase is due to an increased number of interactions caused by the crossing, while the final saturation might be due to a decrease in speeds when approaching capacity. This saturation effect has so far only been found in a small number of empirical studies, so this study lends further support to it. Furthermore, it provides a complete picture of the number of conflicts in relationship to the traffic state in general. 

There are differences between the simulation results and the empirical observations. While $n \propto q_1 q_2$ for the simulation results, and the number of crashes derived from the number of conflicts must follow a Poisson distribution, real data have $N \propto Q_1^{\beta_1} Q_2^{\beta_2}$ with exponents $\beta_i < 1$ and they follow a negative binomial distribution. Therefore, an observation model has been investigated to try to understand how to do the transition from the microscopic variables $(n,q_i)$ to the macroscopic ones $(N, Q_i)$. It turns out that the simplest explanation we have found so far is to assume that the ADT-flows $Q_i$ that enter into the empirical analyses are strongly disturbed by a number of processes. E.\,g., as the example of one loop detector in Fig.~\ref{fig:TE370} shows, even what is associated with one $Q_i$-value carries considerable variance. Furthermore, in reality the process how the $Q_i$ are measured (or estimated) is prone to errors.  
A second problem is with the modelling in typical traffic safety analyses; they are too strongly bound to equations where a GLM can be used. Eq.~(\ref{eq:theory}) is not of this type, and the saturation effects seen in the simulation results are also not very well described by the power-law approach of Eq.~(\ref{eq:glm}). The approach here shows that at least for small traffic flows a mechanistic explanation is possible that we deem better than the simple regression used in most empirical studies. 

Note that there is an additional twist in the transformation from the simulation variables to the empirical ones: Since the $Q_i$ are average values, they do a poor job in sampling maximum values -- the $Q_i$ will never reach the areas where there are strong deviations from the low-flow behavior. However, the $N$ do sample from the whole area of flow values that make up $Q_i$, further muddling the picture. The analyses where saturation have been seen work in fact with short-term flows (e.\,g.~hourly values), and not with the ADT-values of all the other analyses. Which is a nice explanation why some studies do in fact see saturation, while others do not.

Finally, as pointed out already, the crash probability (named $\alpha$ in the description above) itself could be a function of the traffic state and transform the relationships found here by simulation: One effect is when speeds of the vehicles decrease, then at least the severity of the crashes decreases, and one may argue that slower speeds give drivers a longer time-span to react, so the number of crashes might go down with speed as well. This paper has not investigated this avenue, which might be better left to human factors experts, and it will come on top of the simple model explored here.

Clearly, it might be interesting to see whether the approach here holds also for different intersection organizations and for more complicated models, which will be the topic of future work. In addition, the approach here makes a prediction about the ratio between rear-end and crossing conflicts as a function of the traffic flow, which is in principle testable with real data.

\section*{CRediT author statement}

{\bf Andreas Leich:} Conceptualization, Investigation, Methodology, Writing- Reviewing and Editing {\bf Ronald Nippold:} Methodology, Writing- Reviewing and Editing {\bf Andreas Schadschneider:} Conceptualization, Investigation , Writing- Reviewing and Editing {\bf Peter Wagner:} Conceptualization, Methodology, Software, Writing- Original draft preparation, Writing- Reviewing and Editing

\section*{Declaration of interests}
 
The authors declare that they have no known competing financial interests or personal relationships that could have appeared to influence the work reported in this paper.

\bibliographystyle{elsarticle-num} 
\bibliography{TS4X}      

\begin{thebibliography}{10}
\expandafter\ifx\csname url\endcsname\relax
  \def\url#1{\texttt{#1}}\fi
\expandafter\ifx\csname urlprefix\endcsname\relax\def\urlprefix{URL }\fi
\expandafter\ifx\csname href\endcsname\relax
  \def\href#1#2{#2} \def\path#1{#1}\fi

\bibitem{Veh1937}
A.~Veh, Improvements to reduce traffic accidents, Proceedings of the ASCE
  (1937) 1775--1785Meeting of the Highway Division: New York, NY, USA.

\bibitem{Hauer2004}
E.~Hauer, Statistical {R}oad {S}afety {M}odeling, Transportation Research
  Record 1897 (2004) 81--87.
\newblock \href {https://doi.org/10.3141/1897-11} {\path{doi:10.3141/1897-11}}.

\bibitem{LordMannering2010}
D.~Lord, F.~Mannering, The statistical analysis of crash-frequency data: A
  review and assessment of methodological alternatives, Transportation Research
  Part A: Policy and Practice 44 (2010) 291–305.
\newblock \href {https://doi.org/10.1016/j.tra.2010.02.001}
  {\path{doi:10.1016/j.tra.2010.02.001}}.

\bibitem{KononovEtAl2011}
J.~Kononov, C.~Lyon, B.~K. Allery, Relation of flow, speed, and density of
  urban freeways to functional form of a safety performance function,
  Transportation Research Record 2236~(1) (2011) 11--19.
\newblock \href {https://doi.org/10.3141/2236-02} {\path{doi:10.3141/2236-02}}.

\bibitem{Hoye:Hesjevoll:2020}
A.~K. Høye, I.~S. Hesjevoll,
  \href{http://www.sciencedirect.com/science/article/pii/S0001457520307934}{Traffic
  volume and crashes and how crash and road characteristics affect their
  relationship – a meta-analysis}, Accident Analysis \& Prevention 145 (2020)
  105668.
\newblock \href {https://doi.org/https://doi.org/10.1016/j.aap.2020.105668}
  {\path{doi:https://doi.org/10.1016/j.aap.2020.105668}}.
\newline\urlprefix\url{http://www.sciencedirect.com/science/article/pii/S0001457520307934}

\bibitem{Nagatani:1993}
T.~Nagatani, Effect of traffic accident on jamming transition in traffic-flow
  model, J. Phys. A 26 (1993) L1015.
\newblock \href {https://doi.org/10.1088/0305-4470/26/19/008}
  {\path{doi:10.1088/0305-4470/26/19/008}}.

\bibitem{Boccaraetal:1997}
N.~Boccara, H.~Fuks, Q.~Zeng, Car accidents and number of stopped cars due to
  road blockage on a one-lane highway, J. Phys. A 30 (1997) 3329.
\newblock \href {https://doi.org/10.1088/0305-4470/30/10/012}
  {\path{doi:10.1088/0305-4470/30/10/012}}.

\bibitem{Huang:1998}
D.~Huang, Exact results for car accidents in a traffic model, J. Phys. A 31
  (1998) 6167.
\newblock \href {https://doi.org/10.1088/0305-4470/31/29/008}
  {\path{doi:10.1088/0305-4470/31/29/008}}.

\bibitem{Nagel1992}
K.~Nagel, M.~Schreckenberg, A cellular automaton model for freeway traffic,
  Journal de Physique I France 2 (1992) 2221 -- 2229.
\newblock \href {https://doi.org/10.1051/jp1:1992277}
  {\path{doi:10.1051/jp1:1992277}}.

\bibitem{Chowdhury-Review:2000}
D.~Chowdhury, L.~Santen, A.~Schadschneider, Statistical physics of vehicular
  traffic and some related systems, Physics Reports 329~(4--6) (2000) 199--329.

\bibitem{SchadschneiderChNi}
A.~Schadschneider, D.~Chowdhury, K.~Nishinari, Stochastic Transport in Complex
  Systems, Elsevier.
\newblock \href {https://doi.org/10.1016/C2009-0-16900-3}
  {\path{doi:10.1016/C2009-0-16900-3}}.

\bibitem{Ito:Nishinari:2014}
H.~Ito, K.~Nishinari, Totally asymmetric simple exclusion process with a
  time-dependent boundary: Interaction between vehicles and pedestrians at
  intersections, Phys. Rev. E 89 (2014) 042813.
\newblock \href {https://doi.org/10.1103/PhysRevE.89.042813}
  {\path{doi:10.1103/PhysRevE.89.042813}}.

\bibitem{Marzoug:etal:2018}
R.~Marzoug, N.~Lakouari, O.~Oubram, H.~Ez-Zahraouy, A.~Khallouk,
  M.~Lim{\'{o}}n-Mendoza, J.~G. Vera-Dimas, Impact of traffic lights on car
  accidents at intersections, International Journal of Modern Physics C 29~(12)
  (2018) 1850121.
\newblock \href {https://doi.org/10.1142/s0129183118501218}
  {\path{doi:10.1142/s0129183118501218}}.

\bibitem{Marzoug:etal:2021}
R.~Marzoug, O.~Bamaarouf, N.~Lakouari, B.~Tellez, M.~Tellez, O.Oubram, Traffic
  intersection characteristics with accidents and evacuation of damaged cars,
  Physica A 561 (2021) 125217.
\newblock \href {https://doi.org/10.1016/j.physa.2020.125217}
  {\path{doi:10.1016/j.physa.2020.125217}}.

\bibitem{Janvier:Saunier:2020}
L.~N. Janvier, N.~Saunier, An open-source minimal micro-simulation tool for
  safety analysis, in: Transportation Research Board, 2020, pp. 1--19.

\bibitem{Akriti:Gupta:2021}
A.~Jindal, A.~K. Gupta, Exclusion process on two intersecting lanes with
  constrained resources: Symmetry breaking and shock dynamics, Phys. Rev. E 104
  (2021) 014138.
\newblock \href {https://doi.org/10.1103/PhysRevE.104.014138}
  {\path{doi:10.1103/PhysRevE.104.014138}}.

\bibitem{HSM2010}
J.~A. Bonneson, et~al., Highway Safety Manual, American Association of State
  Highway and Transportation Officials, 2010.

\bibitem{WaPetAl:2021}
P.~Wagner, R.~Hoffmann, A.~Leich, Observations on the relationship between
  crash frequency and traffic flow, Safety 7~(1) (2021) 3.
\newblock \href {https://doi.org/10.3390/safety7010003}
  {\path{doi:10.3390/safety7010003}}.

\bibitem{HughesEtAl2015}
B.~Hughes, S.~Newstead, A.~Anund, C.~Shu, T.~Falkmer,
  \href{http://www.sciencedirect.com/science/article/pii/S0001457514001766}{A
  review of models relevant to road safety}, Accident Analysis \& Prevention 74
  (2015) 250 -- 270.
\newblock \href {https://doi.org/10.1016/j.aap.2014.06.003}
  {\path{doi:10.1016/j.aap.2014.06.003}}.
\newline\urlprefix\url{http://www.sciencedirect.com/science/article/pii/S0001457514001766}

\bibitem{ManneringBook2018}
F.~Mannering, Cross-sectional modelling, in: D.~Lord, S.~Washington (Eds.),
  Safe Mobility -- Challenges, Methodology, and Solutions, Emerald Publishing
  Limited, 2018, pp. 257 -- 277.

\bibitem{Ambros2018}
J.~Ambros, C.~Jurewicz, S.~Turner, M.~Kieć, An international review of
  challenges and opportunities in development and use of crash prediction
  models, European Transport Research Review 10~(2) (2018) 35.
\newblock \href {https://doi.org/10.1186/s12544-018-0307-7}
  {\path{doi:10.1186/s12544-018-0307-7}}.

\bibitem{Leich:2022}
A.~Leich, J.~Fuchs, G.~Srinivas, J.~Niemeijer, P.~Wagner, Traffic safety at
  german roundabouts -- a replication study, Safety 8~(3) (2022).
\newblock \href {https://doi.org/10.3390/safety8030050}
  {\path{doi:10.3390/safety8030050}}.

\bibitem{RetallackOstendorf2020}
A.~Retallack, B.~Ostendorf, {Relationship Between Traffic Volume and Accident
  Frequency at Intersections}, International Journal of Environmental Research
  and Public Health 17~(4) (2020) 1393.
\newblock \href {https://doi.org/10.3390/ijerph17041393}
  {\path{doi:10.3390/ijerph17041393}}.

\bibitem{BerlinLoops}
\href{https://api.viz.berlin.de/daten/verkehrsdetektion}{{Digitale Plattform
  Stadtverkehr Berlin - Verkehrsdetektion}} (2023).
\newline\urlprefix\url{https://api.viz.berlin.de/daten/verkehrsdetektion}

\bibitem{UTD19}
L.~A. Loder, Allister, M.~Menendez, K.~W. Axhausen, Understanding traffic
  capacity of urban networks, Scientific Reports 9~(1) (2019) 16283.
\newblock \href {https://doi.org/https://doi.org/10.1038/s41598-019-51539-5}
  {\path{doi:https://doi.org/10.1038/s41598-019-51539-5}}.

\end{thebibliography}

\end{document}